# Dynamic mode evolution and phase transition of twisted light in nonlinear process

Yan Li,[1,2] Zhi-Yuan Zhou,[1,2,#] Dong-Sheng Ding,[1,2] Wei Zhang,[1,2] Shuai Shi,[1,2]

Bao-Sen Shi,[1,2*] and Guang-Can Guo[1,2]

[1]*Key Laboratory of Quantum Information, University of Science and Technology of China, Hefei, Anhui 230026, China*

[2]*Synergetic Innovation Center of Quantum Information & Quantum Physics, University of Science and Technology of China, Hefei, Anhui 230026, China*

[*] *drshi@ustc.edu.cn*
[#] *zyzhouphy@mail.ustc.edu.cn*

We report on studying the dynamic evolution of orbital angular momentum (OAM) carrying light in sum frequency generation (SFG). Dynamic evolution of the SFG beam is studied in different cases, in which the two pump beams are either a single OAM mode or OAM superposition mode. For the cases when the two pumps are both in superposition mode, two kinds of spatial patterns and evolution behaviors are observed, one set of spatial pattern is non-diffractive and unchanged with propagating. In addition, SFG of two pump beams with opposite OAM will evolve into a kind of quasi-Gaussian mode. These observations show that the pumps' phases are coherently transferred to SFG beam in the conversion process. Our findings give insights to physical picture of OAM light transformation in nonlinear processes, which can be used in OAM mode engineering and de-multiplexing of different OAM modes.

Recent years have seen vast progress in the generation and detection of structured light, with potential applications in high capacity optical data storage and optical communications. It is well known that photons can carry both spin and orbital angular momentum (OAM), the spin is associated with the polarizations and the OAM with helical phase structure in the paraxial regime [1]. It is shown by L. Allen et al. that a photon with helical phase form of $e^{il\theta}$ carries $l\hbar$ OAM. The singularity of OAM light has found as a useful tool in many applications [2-8]. OAM modes with different $l$ values are orthogonal, therefore information can be encoded in a higher dimension in contrast to polarization modes. The combination of OAM and polarization encoding has demonstrated to increase the channel capacity of communication systems [8]. In addition, the tolerance of quantum key distribution can be increased by encoding in multi-level OAM basis [9, 10].

Phase-matched nonlinear processes include second order or third order nonlinear processes such as sum frequency generation (SFG), spontaneous parametric down conversion (SPDC), four wave mixing (FWM) and stimulated Raman scattering. Phase matching in the longitude wave vectors results in momentum correlation but it also allows for transfer of transverse phase information. The study of structured light in various nonlinear processes has been a long history.

The conservation, entanglements and interference of OAM light have been studied in second order nonlinear processes such as second harmonic generation (SHG) [11-16], SFG [17-20] and SPDC [21, 22] processes. The transfer of phase structure has also been studied in third order nonlinear processes in FWM [23-25] or higher order nonlinear processes [26]. Recently, the phase structured light has been used for quantum memories in both electromagnetically induced transparency (EIT) [27, 28] and Raman [29] storage schemes. In addition, the spatially dependent EIT is studied and observed both in theory and experiments [30].

For various studies of OAM-carrying light transformation in phase-matched nonlinear processes, the dynamic mode evolution of the generated light by consider it propagation has not been given yet. In this work, the dynamic evolution of 525.5nm structured light generated from SFG of 795nm and 1550nm OAM-carrying light is studied. The two pump beams are either single or superposition of OAM-carrying light. By tracking the mode evolution from the near field to far field, abundant phenomena are observed. The first interesting phenomenon is that SFG of two pump beams with opposite OAM will evolve form a ring structure in the near field to quasi-Gaussian mode in the far field. The second interesting phenomenon is that two sets of patterns and dynamic evolving behaviors are observed when the two pump beams are both in OAM-superposition modes. These two sets of patterns are obtained by changing the relative phase between OAM-superposition modes in one of the pump beam, which shows the transferring of the phase from the pump beam to the SFG beam. The above phenomena have not been studied and observed before. Our findings give clear pictures for OAM mode evolution in SFG process, which can be used for OAM mode engineering and de-multiplexing.

The experimental setup is depicted in figure 1. 795nm OAM-carrying light is generated from spatial light modulator (SLM), the 1550nm OAM-carrying light is generated using vortex phase plate (VPP). The superposition of 1550nm OAM-carrying light is obtained by using a modified Sagnac interferometer. The two pump beams are focused using lenses L1 and L2 separately and combined using dichromatic mirror (DM) before entering the periodically poled KTP crystal (PPKTP, Raicol Crystals, dimensions is 1mm×2mm×10mm, the poling period is 9.375μm). Before measuring the SFG mode using charge coupled device (CCD) camera, the SFG beam is imaged using lens L3 and passed through filters to remove the pump beams.

For SFG of two OAM-carrying beams with OAM index of $m$ and $n$, the SFG beam has the analytical express form of [17]

$$E_{SFG}(r,\alpha,z) = \frac{2[(|m|+|n|-|m+n|)/2]!}{\sqrt{|m|!|n|!}} \frac{(\sqrt{2})^{|m|+|n|}}{w_1^{|m|+1} w_2^{|n|+1}} \frac{i^{m+n+1}}{\lambda z} \exp(-\frac{ik}{2z}r^2 - ikz)\exp(-\frac{k^2 r^2}{4\xi z^2})$$
$$\xi^{-(|m|+|n|+|m+n|)/2-1}\left(\frac{kr}{2z}\right)^{|m+n|} L_{(|m|+|n|-|m+n|)/2}^{|m+n|}(\frac{k^2 r^2}{4\xi z^2})\exp[-i(m+n)\varphi] \quad (1)$$

Where $\lambda$ and $k$ is the wavelength and wave number of the SFG beam respectively. $z$ is the propagation direction; $w_1$ and $w_2$ are two pump beams' waist respectively, and

$$\xi = \frac{1}{w_s^2} + \frac{ik}{2z}, \quad w_s = \frac{w_1 w_2}{\sqrt{w_1^2 + w_2^2}} \quad (2)$$

Here $w_s$ represents SFG beam's waist. The analytical expression equation (1) will give all theoretical simulation results for our experiments in the discussion below.

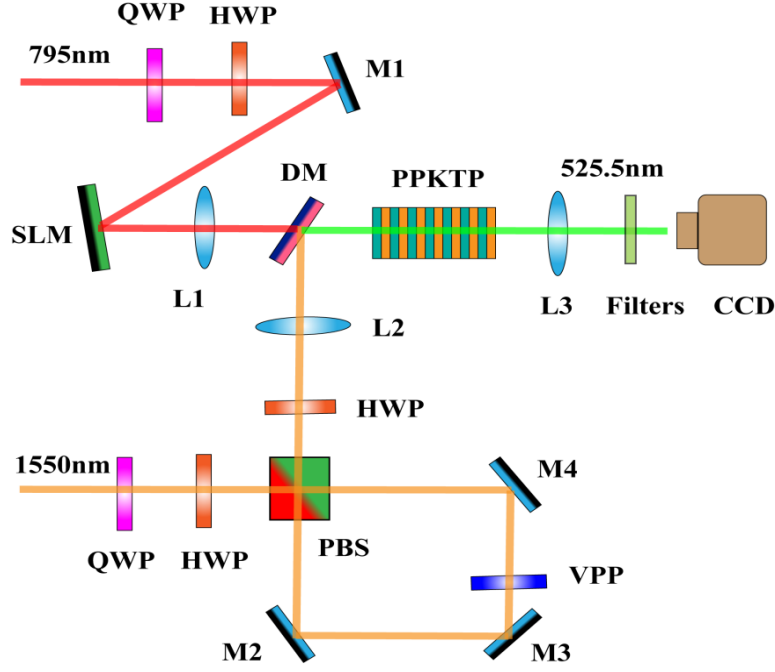

Figure 1. Experimental setup for the experiments. QWP (HWP): quarter (half) wave plate; M1-M4: mirrors; L1-L3: lenses; PBS: polarization beam splitter; DM: dichromatic mirror; VPP: vortex phase plate; SLM: spatial light modulator; PPKTP: periodically poled KTP; CCD: charge coupled device camera.

To investigate the modes evolution dynamics of OAM modes, the case when the two pump beams carry OAM with opposite sign is studied first. The experimental results are showed in figure 2. Images in rows A1-A3 are the experimental results for 795nm pump beam carries OAM index of 1, 2, 3 and 1550nm pump beam carries OAM index of -1, -2, -3 respectively. The images in each row are obtained by moving the CCD camera from near field to far field. In the near filed, the intensity distribution of SFG beam has a single ring shape (images of columns 1, 2 in each row). Then with propagation of the beam away, the intensity of the ring is diming and a central point is appearing, the intensity of the central point becomes brighter and brighter with the propagation distance. Finally in the far field, the SFG beam evolves to a spatial shape with a central bright point and dim concentric rings outside. In the far field approximation ($x \gg kw_s^2/2$), equation (1) simplified to

$$E_{SFG}(r,\alpha,z) = \Gamma \frac{i}{\lambda z} \exp(-ikz) \exp(-\frac{r^2}{w^2}) L_m^0(\frac{r^2}{w^2}) \propto L\tilde{G}_m^0 \quad (3)$$

Where $\Gamma$ is a constant. $w = zw_s/z_R$ and $z_R = kw_s^2/2$ with $w$ and $z_R$ being the spot radius of the SFG beam at $z$ and the Rayleigh range, respectively. $L\tilde{G}_m^0$ is similar to standard Laguerre-Gaussian (LG) mode $LG_m^0$, but they are difference by a factor of 2 in the Laguerre polynomials. This difference makes the intensity in radial direction of the SFG beam decreasing much more rapidly than standard LG mode. Therefore the outer rings in rows A2, A3 can not be seen experimentally. For OAM index of 1, 2, 3, there should be 1, 2, 3 outer rings for the SFG beam in far field. Another property of the far field SFG beam is that most of the power is distributed in the central bright point, the power of the outer rings can be ignored, therefore the mode can be treat as quasi-Gaussian mode. This property will have potential application for OAM mode de-multiplexing. The corresponding theoretical simulation results for rows A1-A3 are showed in rows B1-B3 respectively. The numbers at the top left of each image represent the propagation distance in unit of Rayleigh range ( $z_R = kw_s^2/2$ ).  The simulation results are in well agreement with the experimental observations.

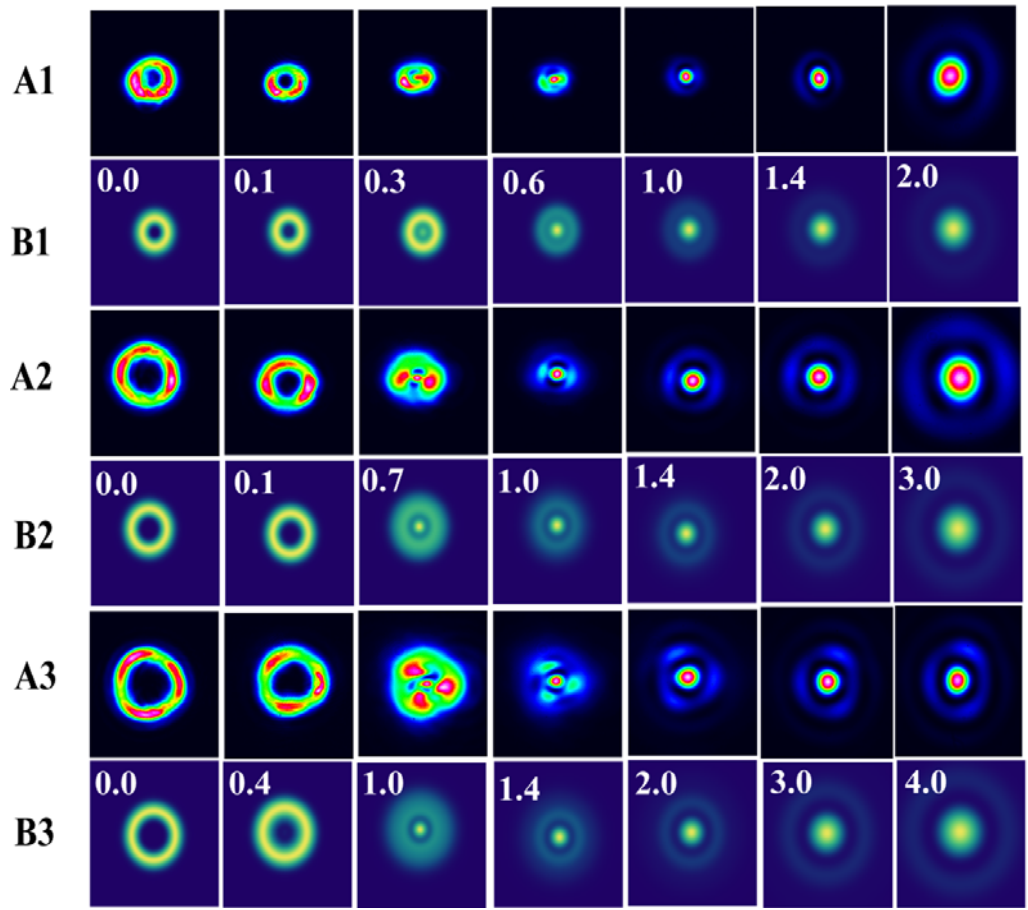

Figure 2. Experimental and simulation results for two pump beams carry the same OAM with opposite sign. Rows A1-A3 for 795nm carries OAM index of 1, 2, 3 and 1550nm carries OAM index of -1, -2, -3, respectively. Images in rows B1-B3 are the corresponding simulation results.

Then we study the mode evolution of the case when 795nm pump beam is OAM superposition mode, and 1550nm pump is a single OAM mode. The experimental and simulation results are showed in figure 3. For rows A1-A3, 795nm pump are in mode $LG_0^l + e^{i\theta} LG_0^{-l} (l=1,2,3)$, 1550nm pump is in mode $LG_0^l (l=1,2,3)$. The results in row A4 are for the modes $LG_0^{l_1} + e^{i\theta} LG_0^{-l_2} (l_1=3, l_2=2)$ and $LG_0^2$ for the 795nm and 1550nm pump beam respectively. Viewing form figure 3, we conclude that SFG beams in the near field is petal-like interference patterns (the number of petals are $2l$ or $l_1 + l_2$). After propagating away, a central point is appearing and it becomes brighter and bright with distance. At the same time the petals are twisted and rotated in the same directions like a windmill. The SFG of two OAM with the same sign will evolve to standard LG mode in the far filed [17], therefore the SFG modes in the far filed in the present case are $LG_0^{2l} + e^{i\theta} L\tilde{G}_l^0$ and $LG_0^{l_1+l_2} + e^{i\theta} L\tilde{G}_{l_2}^0$ for rows A1-A3 and A4, respectively. The far field spatial shapes are interference of a standard LG mode and a non-standard LG mode. The corresponding theoretical results are showed in rows B1-B4, the slight difference between them is result from mode distortion in the conversion process.

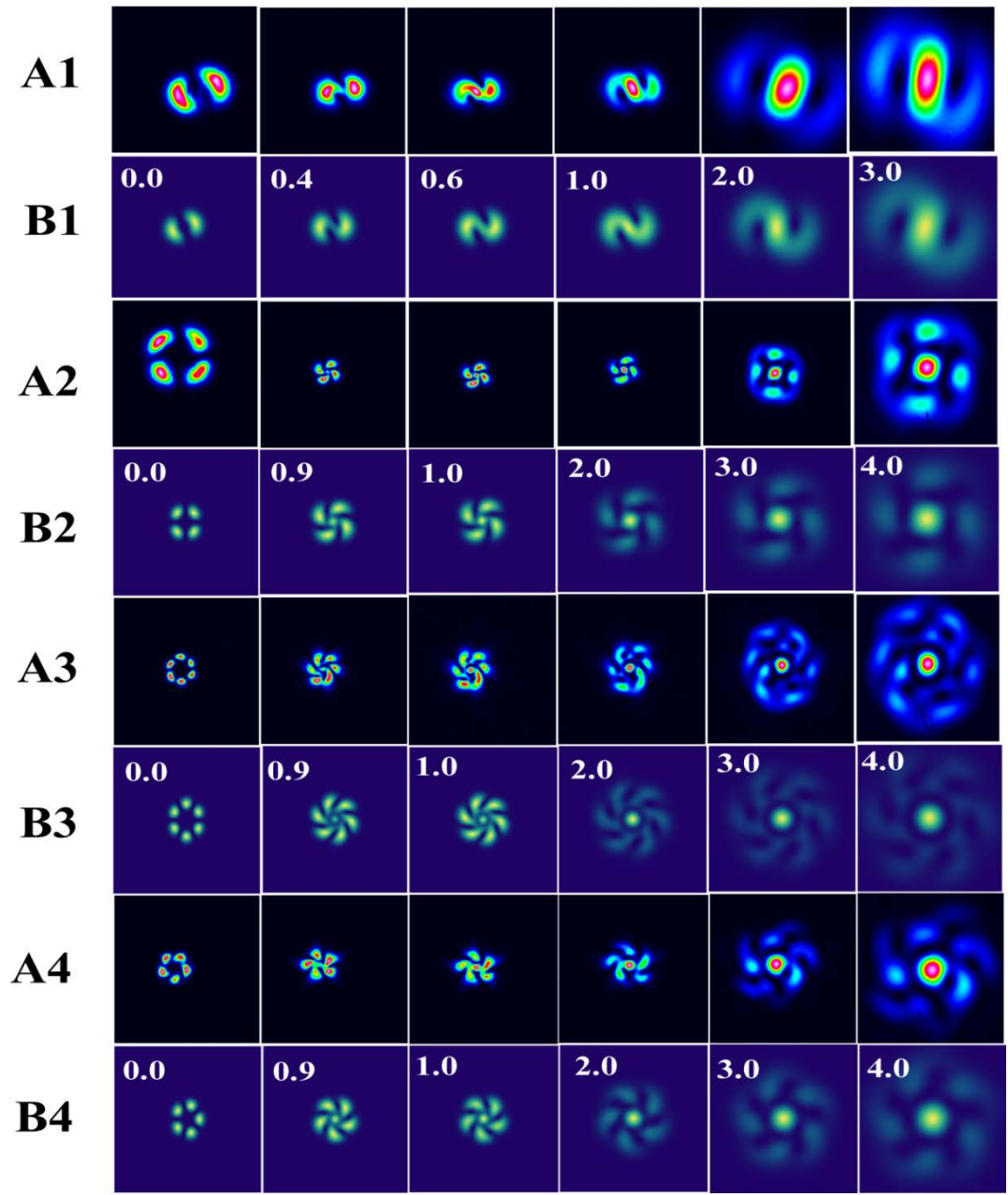

Figure 3. Experimental and simulation results for 795nm pump beam in superposition mode and 1550nm in single OAM mode. Rows A1-A3 for symmetry superposition mode and A4 for asymmetry superposition mode. Rows B1-B4 are the corresponding simulation results.

Finally, we study the case when both the 795nm and 1550nm pump beams are in superposition mode $LG_0^l + e^{i\theta} LG_0^{-l} (l=1,2,3)$. For different relative phase in the superposition, the SFG beam in

far field has the following expression

$$(LG_0^l + e^{i\theta_1} LG_0^{-l}) * (LG_0^l + e^{i\theta_2} LG_0^{-l}) = LG_0^{2l} + e^{i(\theta_1+\theta_2)} LG_0^{-2l} + (e^{i\theta_1} + e^{i\theta_2}) L\tilde{G}_l^0 \quad (4)$$

Equation (4) shows that the far field spatial shape is depend on phase $\theta_1$ and $\theta_2$. There two cases we are concerning: in case I, $\theta_1 = \theta_2$, ie. the two pump beams are in the same mode, the SFG beam becomes $LG_0^{2l} + e^{i2\theta_1} LG_0^{-2l} + 2e^{i\theta_1} L\tilde{G}_l^0$; for case II, $\theta_2 = \theta_1 + \pi$, ie. The two pump beams are orthogonal, the SFG beam simplifies to $LG_0^{2l} - e^{i2\theta_1} LG_0^{-2l}$. These two cases have rather different mode evolution properties and spatial patterns. The experimental and simulation results for case I are showed in figure 4. In the near field, the spatial patterns has petal like structure, the number of petal is $2l$. After propagating away, other petal structure emerges and a central point appears. Finally, the spatial pattern has a bright central point with symmetry dim structures around (see A1-A3). The corresponding simulation results B1-B3 are well matched with the experimental results.

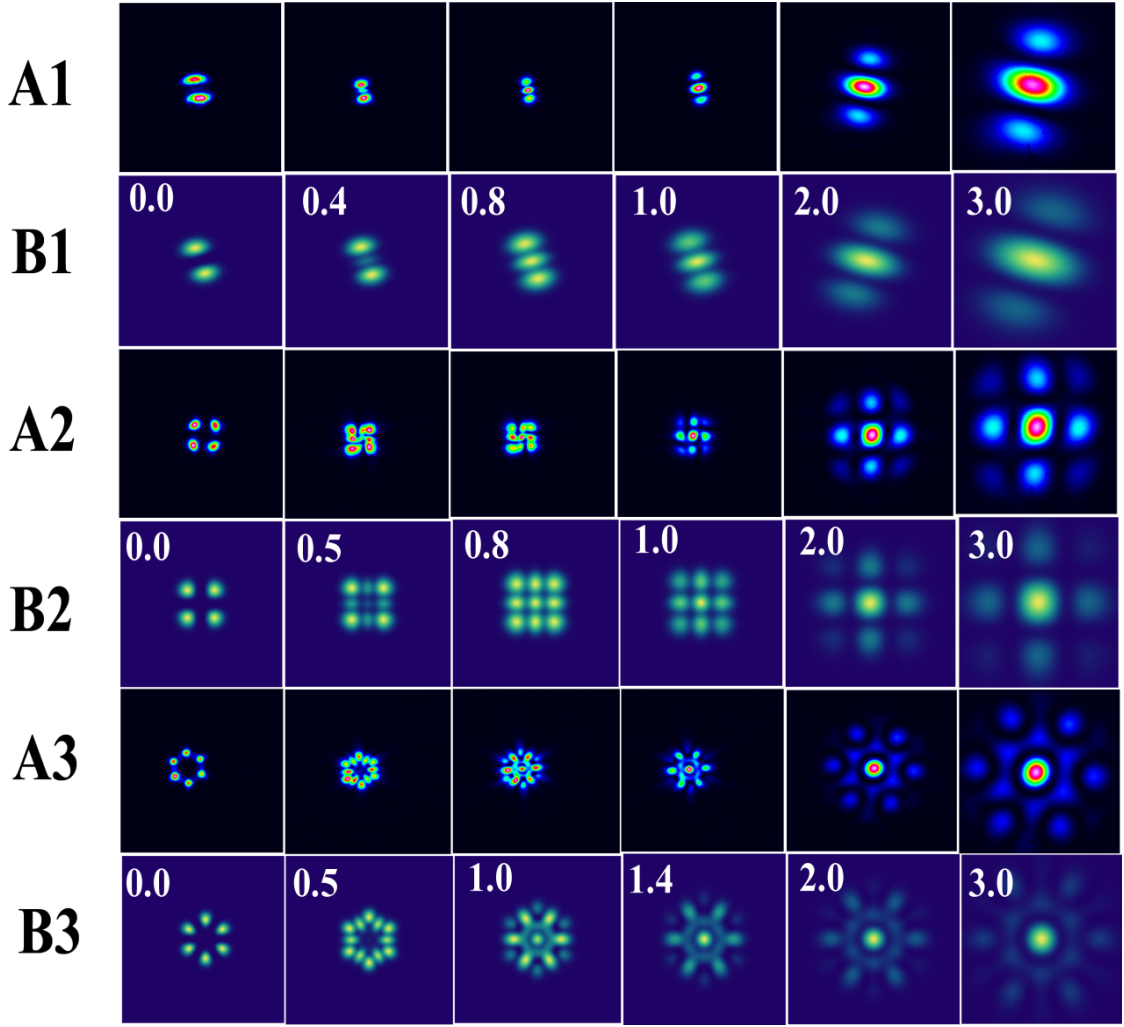

Figure 4. Experimental and simulation results when the two pump in the same superposition mode.

When the two pump beams are in orthogonal superposition modes, the experimental and simulation results are depicted in figure 5. Both the near and far field spatial shapes are petal like structures, the shapes are unchanged with propagating. The numbers of petals are $4l$. This non-diffractive behavior is rather different from case I. The defects of the patterns in the far field are arising from imperfect destructive interference of the cross terms in equation (4).

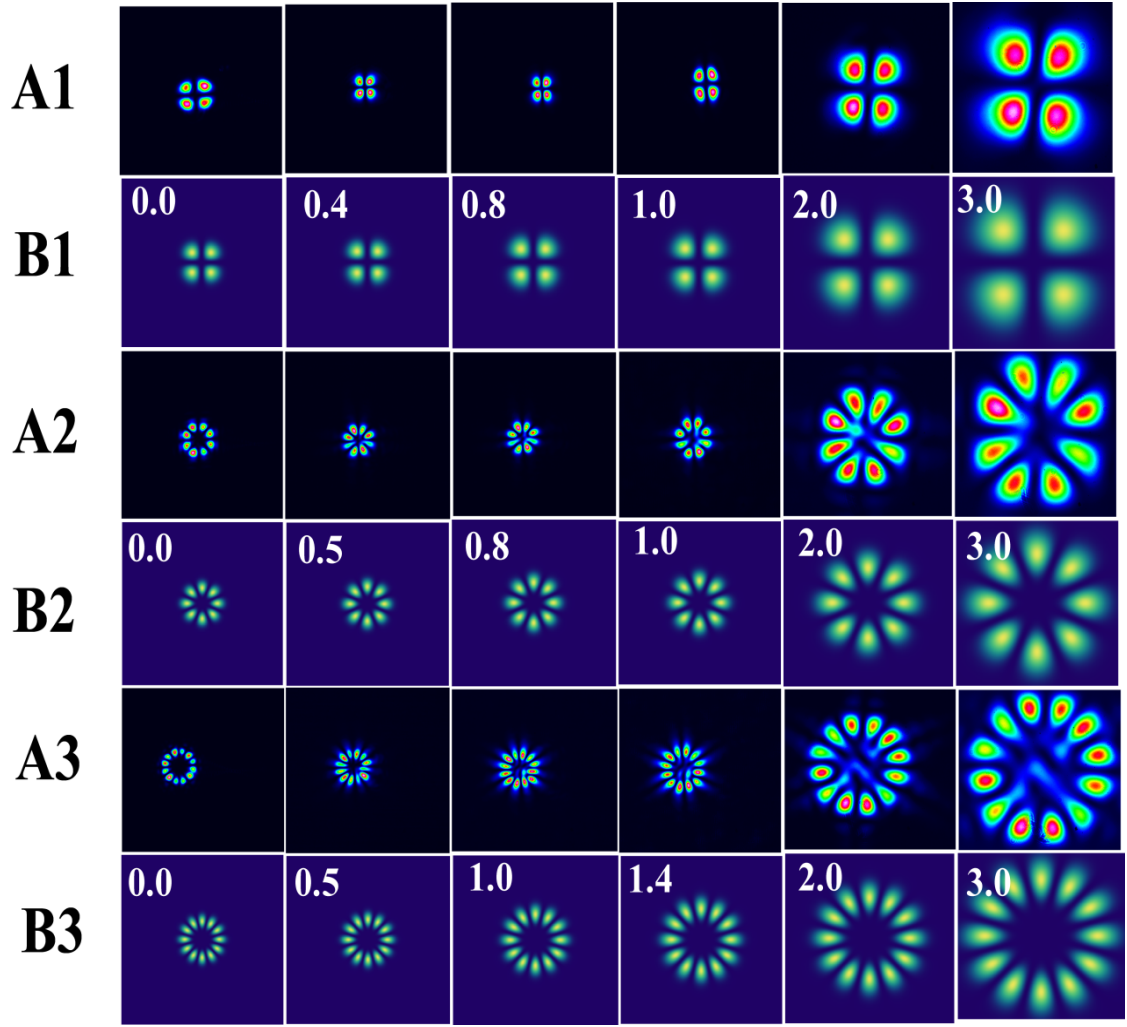

Figure 5. Experimental and simulation results when the two pump in orthogonal superposition mode.

In conclusion, the dynamic mode evolution of OAM in SFG is studied for different cases. A variety of spatial patterns are observed. For two pump beams carry opposite OAM, the SFG beam will evolve from a single ring in the near field to quasi-Gaussian mode in the far field. In the case when one pump beam carries single OAM mode, the other pump carries OAM superposition mode, the SFG beam will evolve from petal like interference pattern in the near field to twisted petals with a central bright point. There are two sets of patterns and propagation behaviors for both pump beams carry OAM superposition mode. For two pump beams with the same OAM superposition mode, the spatial pattern evolves from petal like structure to a central bright point with dim symmetry structure around. While for two pump in orthogonal OAM superposition mode, the spatial pattern is non-diffractive and unchanged along propagation. Our results show that the phases for two pump beams are preserved in the SFG process. The mode evolution dynamics of OAM light of this study is not limited to SFG process, it also applies to other second or third order nonlinear processes. The present study gives insight for OAM mode conversion in SFG and will have potential applications for OAM mode de-multiplexing and engineering.

**Acknowledgements**

This work was supported by the National Fundamental Research Program of China (Grant No. 2011CBA00200), the National Natural Science Foundation of China (Grant Nos. 11174271, 61275115, 61435011) and the Innovation Fund from the Chinese Academy of Science.